\def\Lneq{\mbox{\hspace{0.4em}} {}^{<} \mbox{\hspace{-0.6em}}
  {}_{\sim} \mbox{\hspace{0.4em}}}
\documentstyle[12pt,epsf,subfigure]{ioplppt}
\begin{document}

\begin{center}
  {\LARGE New Universality of Lyapunov Spectra \\in Hamiltonian
    Systems }\\
  \vspace*{1em}
  {\large Yoshiyuki Y. Yamaguchi\\
    Department of Physics, School of Science, Nagoya University,\\
    Nagoya, 464-01, JAPAN}\\
  \vspace*{3em}
\end{center}

\begin{center}
  {\bf\large Abstract}
\end{center}

We show that new universality of Lyapunov spectra $\{\lambda_i\}$
exists in Hamiltonian systems with many degrees of freedom.  
The universality appears in systems which are neither nearly
integrable nor fully chaotic,
and it is different from the one which is obtained in fully chaotic
systems on $1$-dimensional chains as follows.
One is that the universality is found in a finite range of large $i/N$
rather than the whole range, where $N$ is degrees of freedom.
Another is that Lyapunov spectra are not straight,
while fully chaotic systems give straight Lyapunov spectra even 
on the $3$-dimensional simple cubic lattice.
The universality appears when quadratic terms of a potential function
dominate higher terms,
harmonic motions are hence regarded as the base of global motions.

\section{Introduction}
\label{sec:intro}

In the last a few decades Hamiltonian systems with many degrees of
freedom have been numerically investigated by integrating equations of 
motions.
Hamiltonian systems being the foundation of statistical mechanics,
one direction of the investigations is to check ergodic property
originating in Fermi-Pasta-Ulam problem
\cite{fermi55}\cite{saito70}\cite{thirumalai89}. 
Another direction is to study their dynamical properties, which are
transition from nearly integrable systems to stronger chaotic ones
\cite{flach94}\cite{mutschke93},
dynamical properties of phase transition
\cite{antoni95}\cite{yamaguchi96}, 
structure of phase spaces \cite{mutschke93}\cite{yamada87}
and so on.

Here we focus on dynamical properties, in particular, 
universal structures of phase spaces which are not affected by details 
of systems.
One of the structures is self-similar structure which is based on 
Poincar\'e-Birkhoff theorem \cite{lichtenberg92} in nearly integrable
systems. 
Although this theorem is available in systems with two degrees of
freedom, 
the self-similar structure is also supposed in systems with many
degrees of freedom.

Lyapunov spectrum is usually used to study instability along a sample
orbit.
Moreover, it is useful to study structure of phase spaces both in 
dissipative \cite{ikeda86}\cite{nakagawa95} and Hamiltonian
\cite{mutschke93}\cite{yamada87} systems with many degrees of freedom,
since it includes information on all directions in phase space.
Although Lyapunov spectra in a system reveal dynamical properties of
the system,
we are interested further in properties which are not affected by
details of systems.
To detect such properties a useful approach is to find the universal
form of Lyapunov spectra which is obtained in all the systems
and whose cause indicates the properties.
A universal form of Lyapunov spectra $L(i/N)=\lambda_i$ is reported in 
Hamiltonian systems which are $1$-dimensional chains consisting of
nonlinear oscillators \cite{livi87}, where $N$ is degrees of freedom.
This universality gives the straight form for Lyapunov spectra,
namely $L(i/N)=a+b\cdot i/N$,
and the straight form is also obtained with random matrices
\cite{livi87}\cite{eckmann88}.
Consequently, systems having straight Lyapunov spectra are
regarded as fully chaotic ones which have not structure in phase
spaces. 

In this paper, we show new universality of Lyapunov spectra
appearing in systems with moderate strength of chaos
which are neither fully chaotic nor nearly integrable, 
and study the cause of the universality.
Here we define the word ``universality of Lyapunov spectra'' as 
$L(i/N)$ approximately takes the same form in a finite range of large
$i/N$ regardless of total energy and details of systems.
Systems with moderate strength of chaos are interesting
for the following two reasons:
one is that they have some unsolved problems and show interesting
phenomena, for instance, second order phase transition
\cite{yamaguchi97},
and the other is that structure of phase spaces are understood in
systems which are nearly integrable or fully chaotic. 
Models investigated in this paper consist of nonlinear oscillators
with nearest neighbour interactions, 
and each oscillator is on a lattice point of the $3$-dimensional
simple cubic lattice. 

This paper is constructed as follows.
We introduce five models in section \ref{sec:models}. 
They are used to confirm the new universality of Lyapunov
spectra appears in a wide class of Hamiltonian systems.
We show Lyapunov spectra yielded by using random matrices
with temporal $\delta$- or exponential correlations 
in section \ref{sec:RM}.
The Lyapunov spectra are straight even in systems being on
the $3$-dimensional lattices when the systems are fully chaotic.
In sections \ref{sec:universality} and \ref{sec:FPU}, 
we show that Lyapunov spectra for the five systems are not straight
and have universality, and that the systems are neither fully chaotic
nor nearly integrable. 
In section \ref{sec:FPU} 
we show that the new universality appears when quadratic terms
of a potential function, $U_2$, dominate higher terms, $U_4$,
namely $U_2/U_4$ takes large values.
Section \ref{sec:summary} is devoted to summary and discussions.

\section{Models}
\label{sec:models}

We introduce five model Hamiltonians 
each of which represents a system being on the $3$-dimensional simple
cubic lattice with nearest neighbour interactions and periodic
boundary condition.
All Hamiltonians consist of kinetic and potential terms,
\begin{equation}
  H(q,p) = K(p) + U(q), 
  \label{hamiltonians}
\end{equation}
where the kinetic term is
\begin{equation}
  K(p) = \sum^N_{j=1} \frac{1}{2}  p_j^2, 
  \label{kinetic}
\end{equation}
and $N$ is degrees of freedom, 
namely $N=L^3$ where $L$ is the linear size of the lattice.

One of the models is called XY model and expressed as follows
\begin{equation}
  U_{XY}(q) = \sum_{<ij>} [ 1- \cos(q_i-q_j) ], \qquad
  q_j \in [0,2\pi),
  \label{hamiltonian1}
\end{equation}
where the summation $\sum_{<ij>}$ takes over all the pairs of nearest
neighbour lattice points $i$ and $j$. 

The following three systems are expressed as
\begin{equation}
  U(q) = \sum_{<ij>} \frac{1}{2} (q_i-q_j)^2 + \sum^N_{j=1} V(q_j),
  \label{hamiltonian234}
\end{equation}
and they are distinguished by their potential functions $V(q)$.
In Double Well (DW) model
\begin{equation}
  V_{DW}(q) = -\frac{1}{2}q^2 + \frac{1}{4}q^4,
  \label{potential2}
\end{equation}
in Single Well (SW) model
\begin{equation}
  V_{SW}(q) = \frac{1}{2}q^2 + \frac{1}{4}q^4,
  \label{potential3}
\end{equation}
and in Lorentzian (LO) model
\begin{equation}
  V_{LO}(q) = \frac{q^2}{1+q^2}.
  \label{potential4}
\end{equation}

The last one has interactions of FPU-$\beta$ type (3DFPU)
\begin{equation}
  U_{3DFPU} = \sum_{<ij>} \left[ \frac{k}{2} (q_i - q_j)^2
  + \frac{1}{4} (q_i - q_j)^4 \right].
  \label{hamiltonian-FPU}
\end{equation}
This model is used in section \ref{sec:FPU} to determine which term is 
dominant when the new universality appears.

Numerical integrations of Hamiltonian equations of motion,
\begin{equation}
  \frac{\d q_j}{\d t} = \frac{\partial H(q,p)}{\partial p_j}, \quad
  \frac{\d p_j}{\d t} = -\frac{\partial H(q,p)}{\partial q_j}, \quad
  (j=1,2,\cdots,N),
  \label{eq-motion}
\end{equation}
are performed with fourth order symplectic integrator with the fixed
time slice $\Delta t=0.01$. 
Accuracy of total energy is $\Delta E/E \sim O( (\Delta t)^4 )$ where
$\Delta E$ and $E$ are error and an initial value of total energy,
respectively. 

\section{Lyapunov spectra with random matrices}
\label{sec:RM}

As mentioned in the Introduction,
Lyapunov spectra $L(i/N)$ calculated with random matrices are straight
in $1$-dimensional chains.
Lyapunov spectrum is a set of Lyapunov exponents $\{\lambda_i\}\
(i=1,2,\cdots,D)$, where $D$ is dimension of phase space and the
exponents are put in order as $\lambda_i \geq \lambda_{i+1}$.
The summation of $\lambda_i$ up to $n$, $\sum^n_{i=1} \lambda_i$,
indicates linear instability of $n$-dimensional volume element $V_n$
in phase space along a sample orbit. 
Namely, $V_n$ diverges or converges as
\begin{equation}
 V_n(t)\sim\exp[(\lambda_1+\lambda_2+\cdots+\lambda_n)t],
\end{equation}
where $t$ represents time.
Each of Hamiltonian systems with $N$ degrees of freedom has $2N$
Lyapunov exponents which satisfy the following relations induced by
symplectic properties
\begin{equation}
  \lambda_{2N-i+1} = -\lambda_{i} \quad (i=1,2,\cdots,N).
\end{equation}
Hence we have only to observe the first half of Lyapunov spectrum when
we consider Hamiltonian systems.
Details of Lyapunov spectrum are reviewed in Ref.\cite{eckmann85} and
references therein.

The purpose of this section is to show that 
fully chaotic systems have straight Lyapunov spectra even in the
$3$-dimensional simple cubic lattice. 
We describe how we calculate Lyapunov spectra with random matrices,
and then the Lyapunov spectra are shown.

Lyapunov spectra indicate linear instability of a sample orbit,
and they are calculated from linearized equations of motion,
\begin{equation}
  \frac{\d}{\d t}
  \left(
    \begin{array}{c}
      \delta q \\
      \delta p
    \end{array}
  \right)
  =
  \left(
    \begin{array}{cc}
      0 & 1_N \\
      -A(q) & 0
    \end{array}
  \right)
  \left(
    \begin{array}{c}
      \delta q \\
      \delta p
    \end{array}
  \right),
  \label{linearized-eq-motion}
\end{equation}
where $(\delta q, \delta p)$ is a tangent vector,
$1_N$ is the unit matrix of $N\times N$, 
and $(i,j)$ element of the matrix $A(q)$ is
\begin{equation}
  A_{ij}(q) = \frac{\partial^2 U(q)}{\partial q_i \partial q_j}.
\end{equation}
Here we used the form of our Hamiltonians, 
equations (\ref{hamiltonians}) and (\ref{kinetic}).
Temporal evolution of the matrix $A(q)$ is determined by temporal
evolution of $q_j(t)$'s,
and we assume that $q_j(t)$'s are independent random variables with
$\delta$- or exponential correlations, namely
\begin{equation}
  C_{ij}(t) \propto \delta_{ij}\delta(t) \quad
  \mbox{or} \quad \delta_{ij} e^{-\alpha t}.
  \label{delta-exp}
\end{equation}
$C_{ij}(t)$ is the correlation function between $q_i(t)$ and
$q_j(t)$, and it is defined as  
\begin{equation}
  C_{ij}(t) = \overline{\Delta q_i(t) \Delta q_j(0)}
  =\lim_{T\to\infty} \frac{1}{T} \int^T_0 dt' 
  \Delta q_i(t+t') \Delta q_j(t'),
\end{equation}
and
\begin{equation}
  \Delta q_j(t) = q_j(t) - \lim_{T\to\infty} \frac{1}{T} \int^T_0 dt
  q_j(t).
\end{equation}

\begin{figure}
  \begin{center}
    \leavevmode
    \epsfxsize=6.5cm
    \subfigure[XY]
    {\epsffile{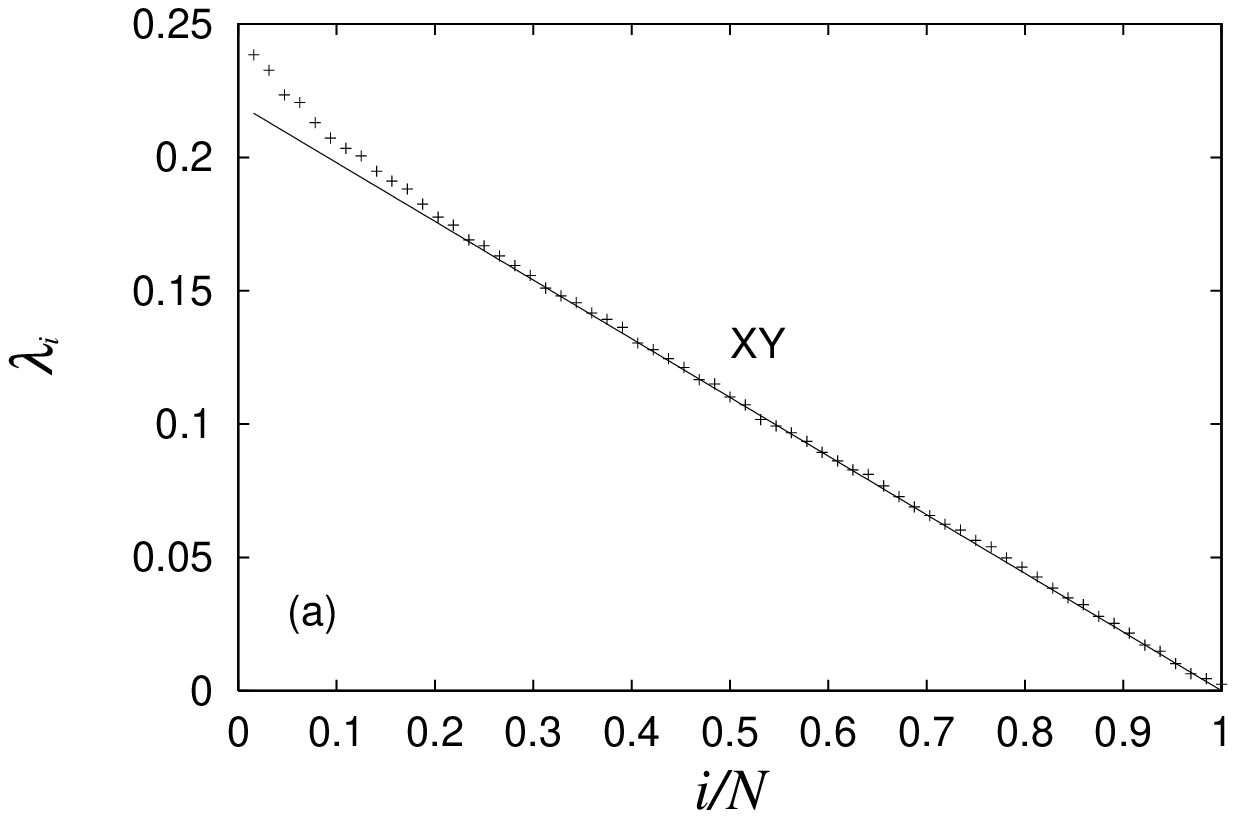}}
    \epsfxsize=6.5cm
    \subfigure[DW and SW]
    {\epsffile{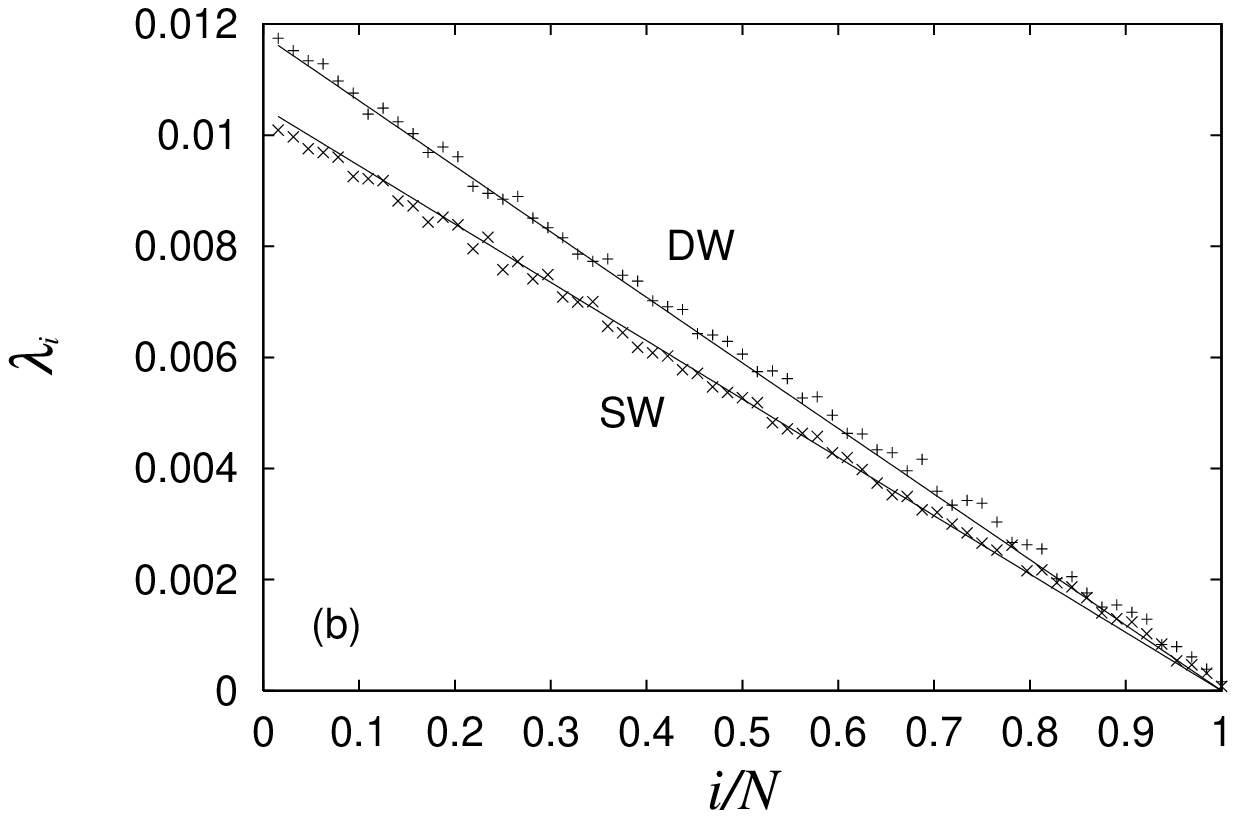}}
    \caption{Lyapunov spectra yielded by using random
      matrices with $\delta$-correlation for XY, DW and SW models.
      (a) XY model. (b) DW ($\diamondsuit$) and SW ($+$) models.
      They are straight even in systems on 3-dimensional
      lattices as they are in 1-dimensional chains.
      Random variables follow uniform distributions.
      The straight solid lines are guides for eyes.}
    \end{center}
    \label{fig:RM}
\end{figure}

We show Lyapunov spectra in figure \ref{fig:RM} which are calculated
with random matrices having the $\delta$-correlation, 
and we find that the Lyapunov spectra are straight
as they are in $1$-dimensional chains.
We take random variables from the uniform distribution, 
and ranges of $q_j(t)$'s are $[-\pi,\pi)$ for XY model, and $[-3,3]$
for DW and SW models.

\begin{figure}
  \begin{center}
    \leavevmode
    \epsfxsize=6.5cm
    \epsffile{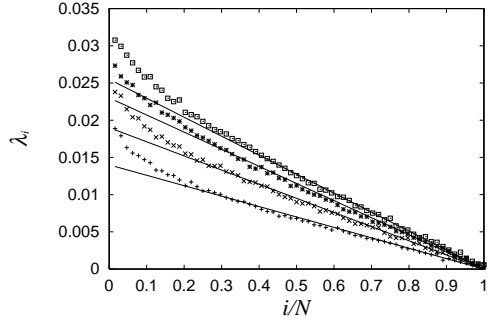}
    \caption{Lyapunov spectra yielded by using random matrices with
      exponential correlation for DW.
      $\alpha$, the reciprocal number of correlation time,
      is $0.4, 0.6, 0.8$ and $1.0$ from lower to upper.
      Spectra show straight behaviour as figure \protect\ref{fig:RM}
      in the region of $0.2 \Lneq i/N \leq 1$.
      The straight solid lines are guides for eyes.}
    \end{center}
    \label{fig:RM-exp}
\end{figure}

Next we change $\delta$-correlation into exponential,
and we set $\alpha=0.4, 0.6, 0.8$ and $1.0$ where $\alpha$ is the
reciprocal number of correlation time of $q_j(t)$ (see equation
(\ref{delta-exp})). 
Figure \ref{fig:RM-exp} shows Lyapunov spectra obtained by using random 
matrices with exponential correlation,
and the Lyapunov spectra are also straight in the region of 
$0.2 \Lneq i/N \leq 1$.
Consequently, we suppose finite correlation time does not affect the
straightness of Lyapunov spectra.

Random matrices with $\delta$- or exponential correlation
yield straight Lyapunov spectra,
and fully chaotic systems hence give the straightness
even in the $3$-dimensional simple cubic lattice.
This result is used later to distinguish our systems evolved by
Hamiltonian equations of motion from fully chaotic ones.

\section{New Universality of Lyapunov spectra}
\label{sec:universality}

In this section, we show that new universality of Lyapunov
spectra exists in systems which are neither fully chaotic nor nearly
integrable and which are in the thermodynamic limit ($N\to\infty$)
through giving the following four results.
(i) The degrees of freedom $N=4^3$ is high enough to reach the
thermodynamic limit for Lyapunov spectra.
(ii) Forms of Lyapunov spectra are invariant with respect to energy 
in each of the four models which are XY, DW, SW and LO
in a finite range of large $i/N$.
(iii) The invariant forms of the four systems are in well agreement,
and they are not approximated by the straight line.
Here we conclude that new universality of Lyapunov spectra is
found.
(iv) Appearance of the new universality is not limited in nearly
integrable systems.

\begin{figure}
  \begin{center}
    \leavevmode
    \epsfxsize=6.5cm
    \subfigure[Non-scaled]
    {\epsffile{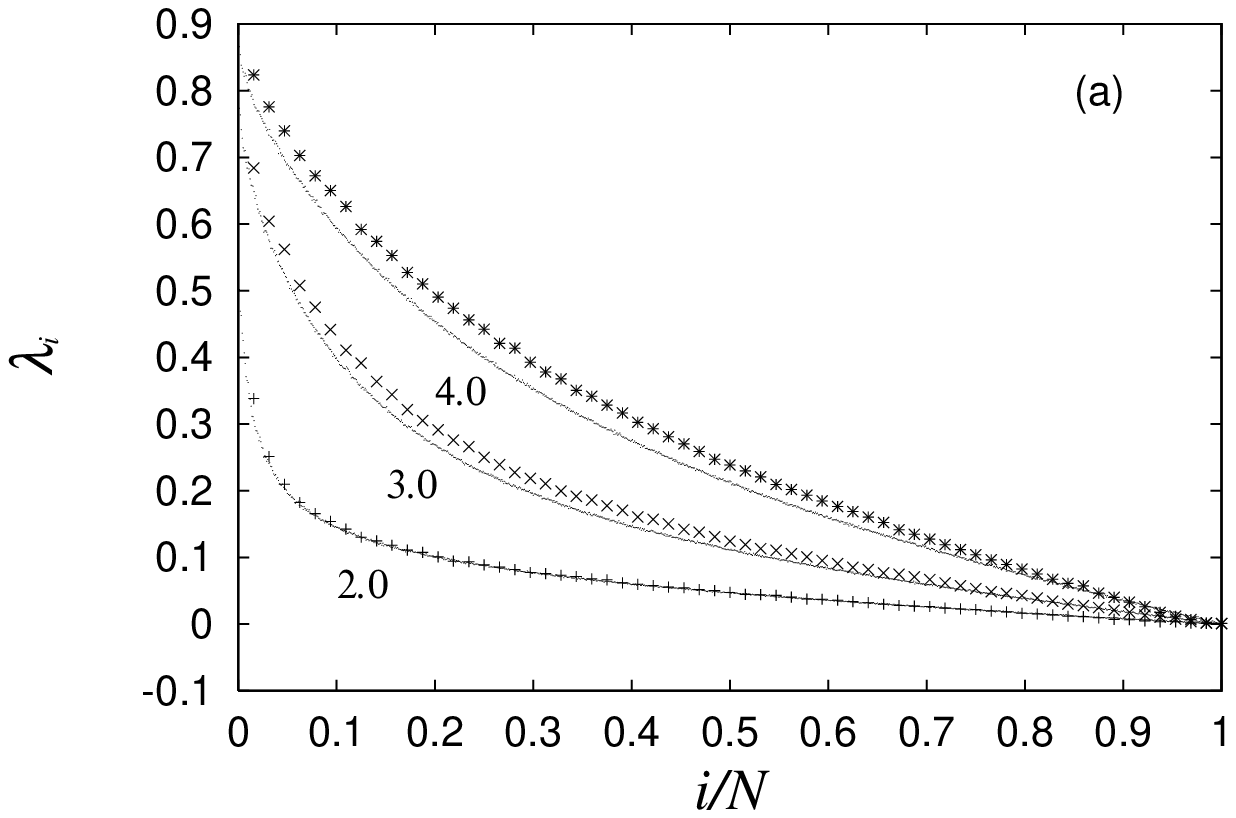}}
    \epsfxsize=6.5cm
    \subfigure[Scaled]
    {\epsffile{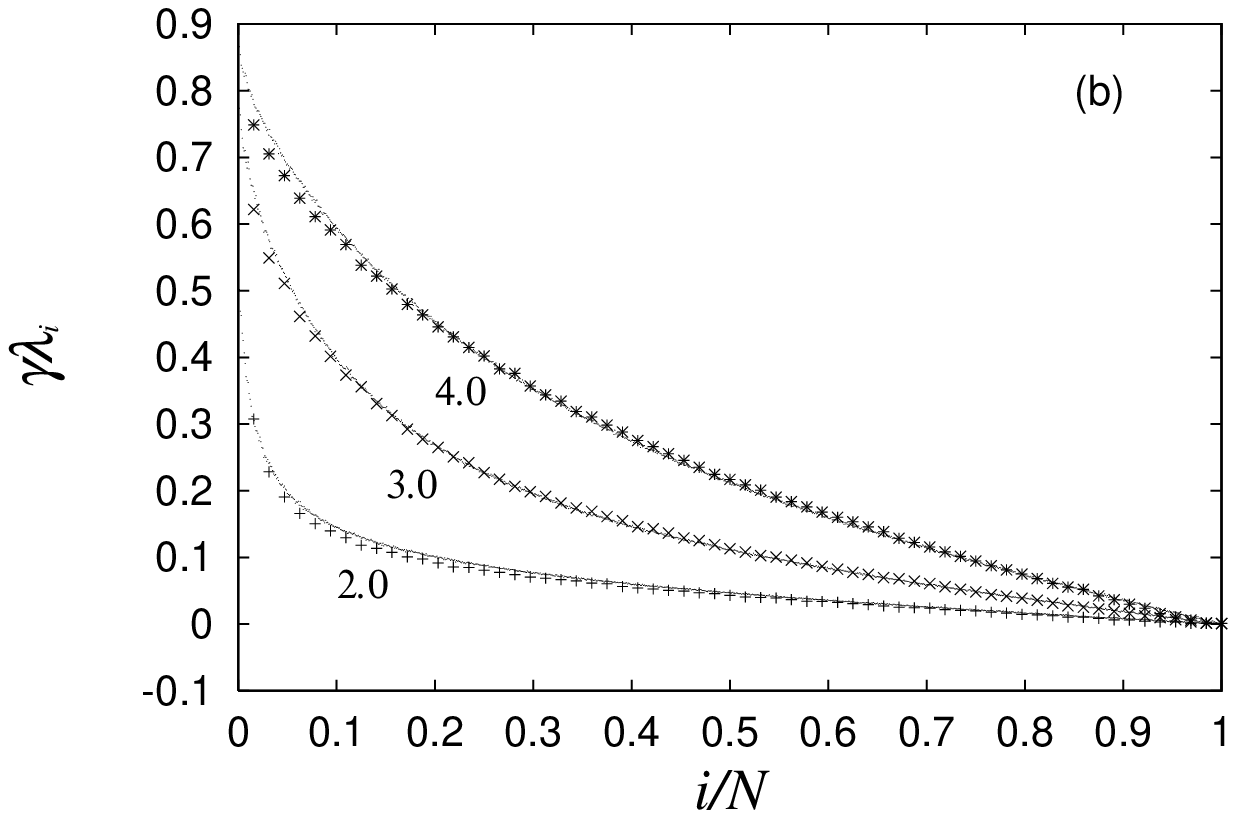}}
    \caption{Dependence on degrees of freedom $N$. Points and dots
      are Lyapunov spectra for $N=4^3$ and $10^3$ respectively.
      Numbers in the figure represent values of energy density $E/N$.
      (a) Non-scaled. (b) Vertical axis is scaled for $N=4^3$ and the
      scale factor $\gamma=1/1.1$ for all values of energy. 
      In each energy, scaled Lyapunov spectrum for $N=4^3$
      is in well agreement with one for $N=10^3$.}
    \end{center}
    \label{fig:N-dep}
\end{figure}

Lyapunov spectra for XY model are shown in figure \ref{fig:N-dep}(a)
in which points and dots represent that the degrees of freedom are
$N=4^3$ and $10^3$ respectively.
Numbers in the figure are values of energy density $E/N$.
The Lyapunov spectra do not correspond with each other
even though they have the same energy density.

Let us note that the form of Lyapunov spectrum $L(i/N)$ is determined
by ratios between Lyapunov exponents $\lambda_i$ rather than their
absolute values.
Accordingly, we may uniformly scale Lyapunov spectrum up or down
from $L(i/N)$ to $\gamma L(i/N)$,
where $\gamma$ is arbitrarily picked for each spectrum.
To multiple $L(i/N)$ by $\gamma$ corresponds
to change time scale from $t$ into $t/\gamma$.
Scaled Lyapunov spectra are shown in figure \ref{fig:N-dep}(b),
and their forms holds regardless of degrees of freedom,
where scale factors are $\gamma=1/1.1$ and $1.0$ for each spectrum in
$N=4^3$ and $10^3$ respectively.
Since the thermodynamic limit of Lyapunov
spectra \cite{livi86}\cite{ruelle82} exists,
we suppose that the system reaches the thermodynamic limit even
$N=4^3$ concerning Lyapunov spectra.
We therefore set $N=4^3$, and use scaled Lyapunov spectra without
comment hereafter.

\begin{figure}
  \begin{center}
    \leavevmode
    \epsfxsize=6.5cm
    \subfigure[XY]
    {\epsffile{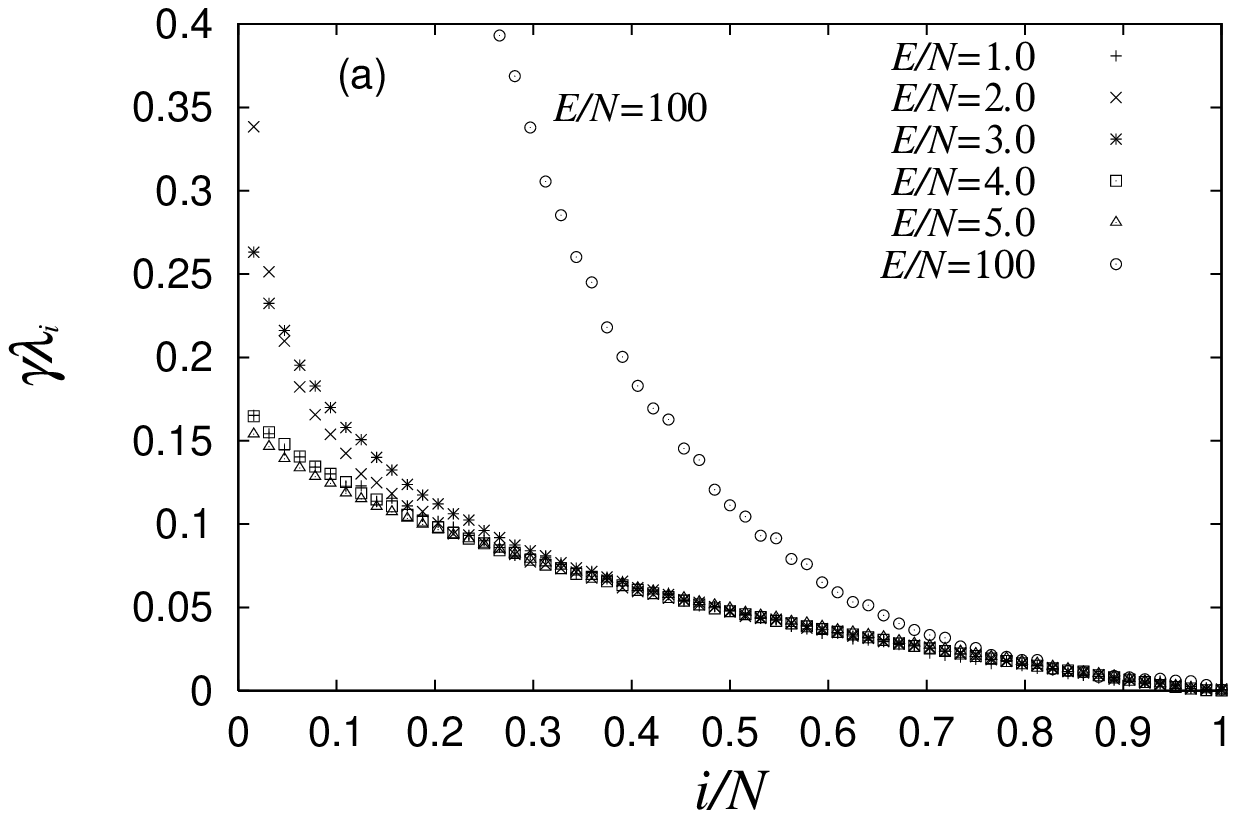}}
    \epsfxsize=6.5cm
    \subfigure[DW]
    {\epsffile{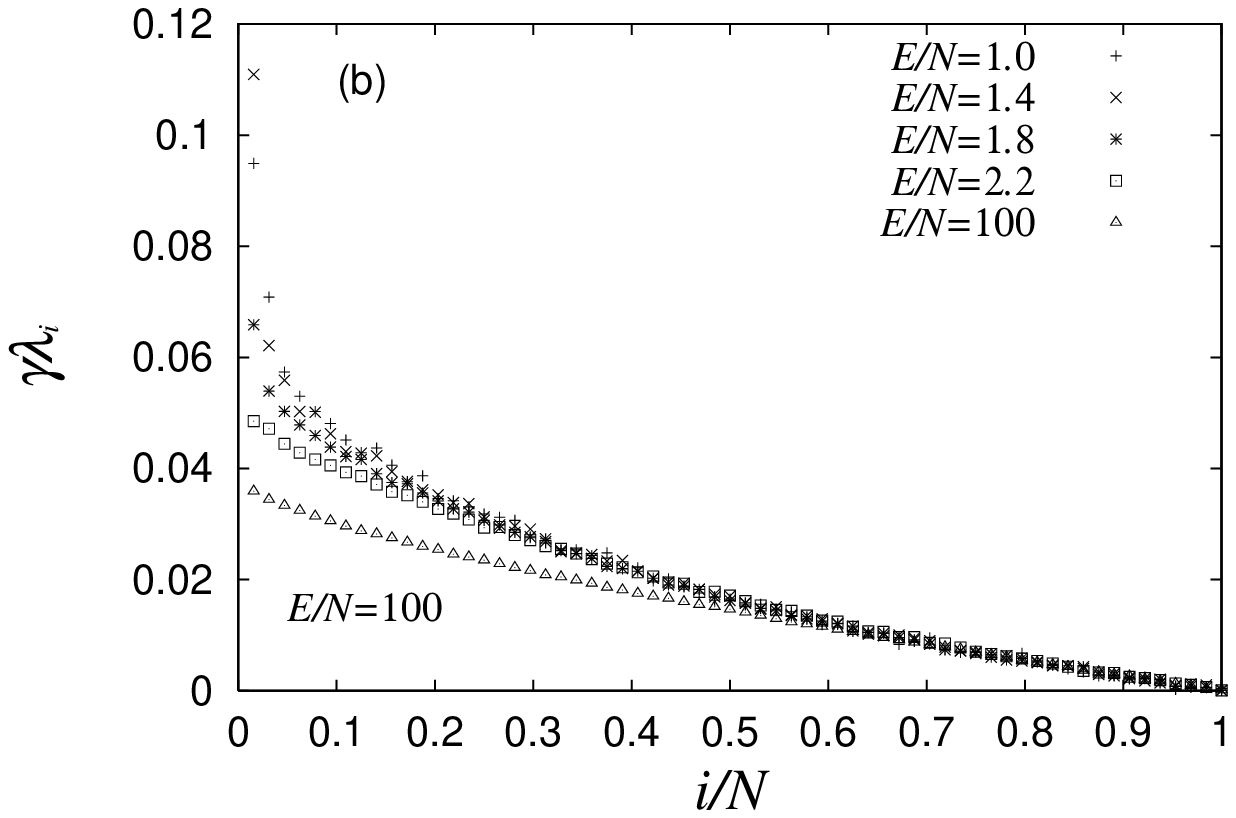}}\\
    \epsfxsize=6.5cm
    \subfigure[SW]
    {\epsffile{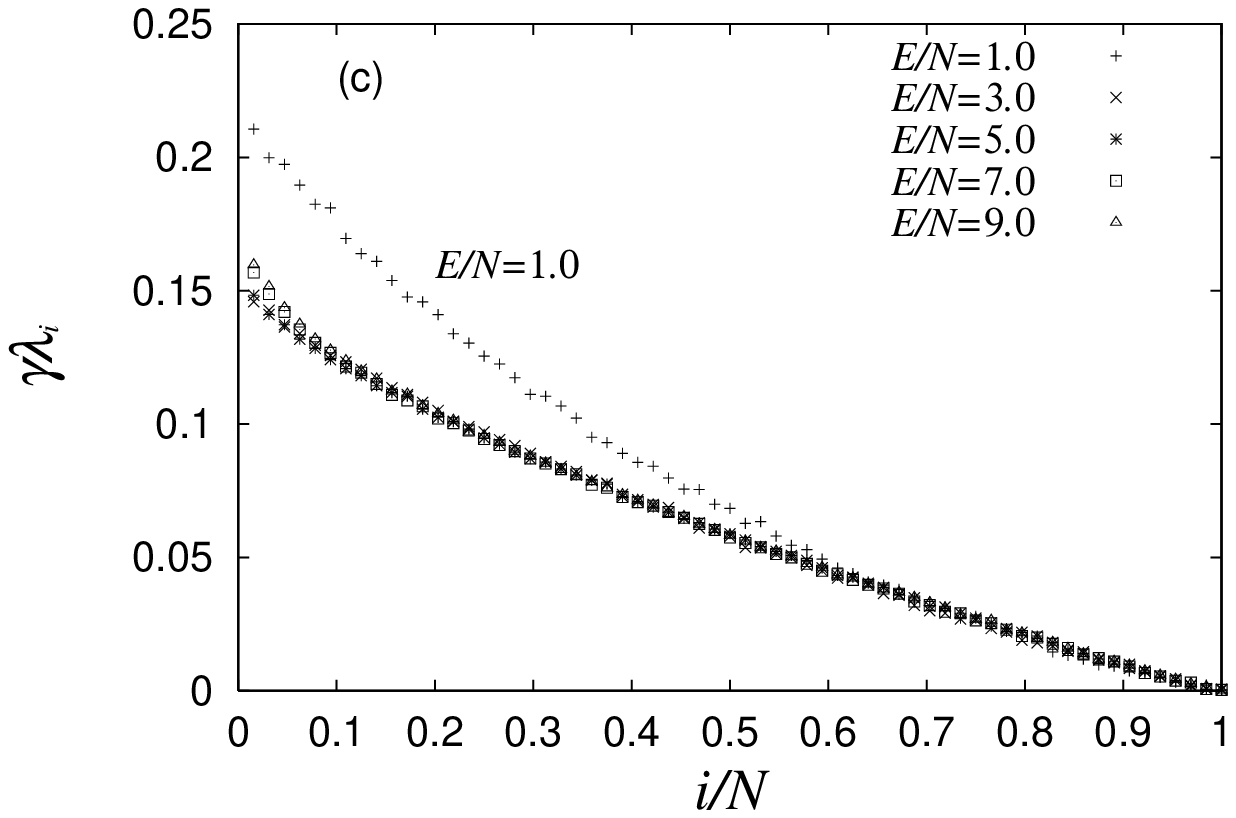}}
    \epsfxsize=6.5cm
    \subfigure[LO]
    {\epsffile{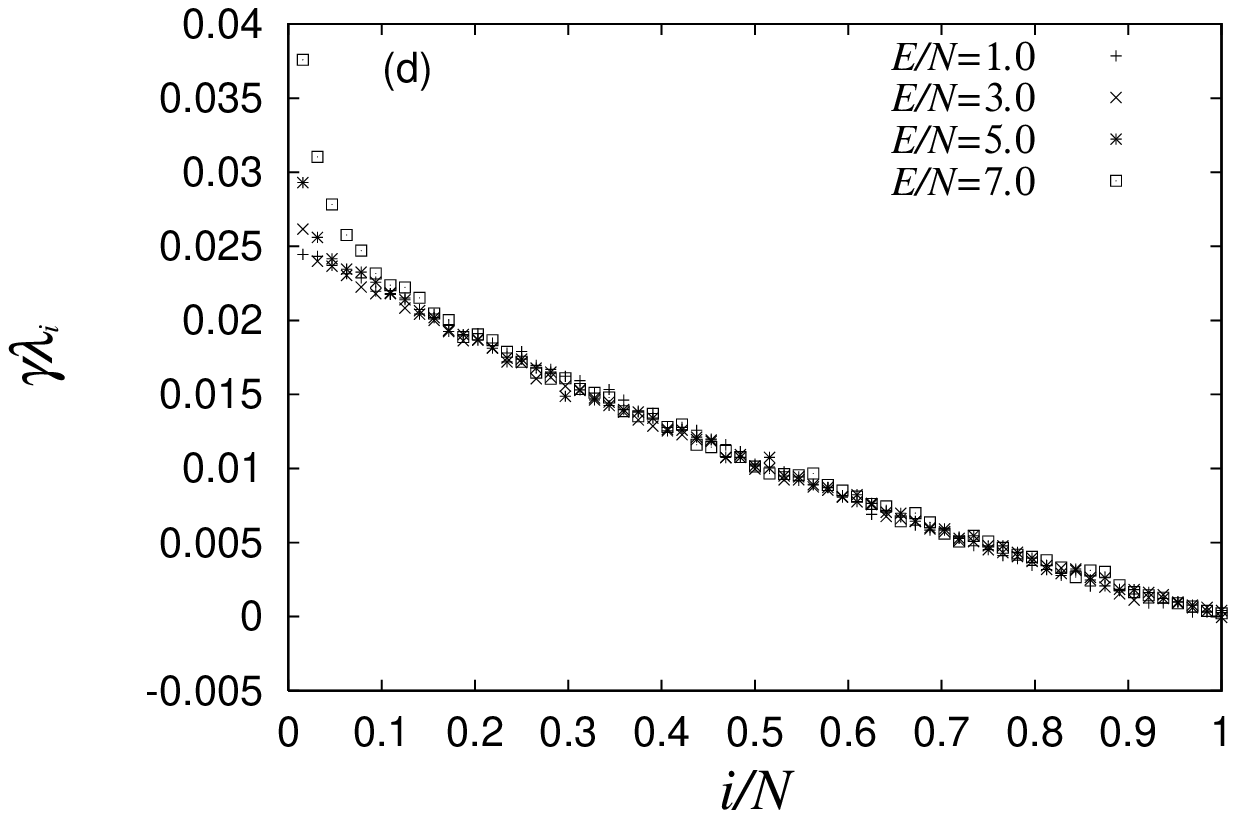}}
    \caption{Lyapunov spectra for various values of energy in four
      Hamiltonian systems.
      (a) XY model. (b) DW model. (c) SW model. (d) LO model. 
      Forms of Lyapunov spectra are the same in a range of large
      $i/N$ in middle energy regime in each model.
      Scale factor $\gamma$ and the range of $i/N$ giving 
      the invariant form are arranged in tables
      \ref{tab:for-fig:scaled} and \ref{tab:range}.}
    \end{center}
    \label{fig:scaled}
\end{figure}

Invariance of $L(i/N)$ with respect to energy is shown in figure
\ref{fig:scaled} for the four models.
In each of the models, 
Lyapunov spectra are in well agreement among various values of energy
in a range of large $i/N$ in middle energy regime.
Scale factor $\gamma$ and the ranges of $i/N$ where the invariance
appears are arranged in tables \ref{tab:for-fig:scaled} and
\ref{tab:range}, respectively.
We remark that the invariance breaks or is strictly limited in a
narrow range of $i/N$ when energy is too high or low,
that are $E/N=100$ in XY and DW, and $E/N=1.0$ in SW.

\begin{figure}
  \begin{center}
    \leavevmode
    \epsfxsize=6.5cm
    \epsffile{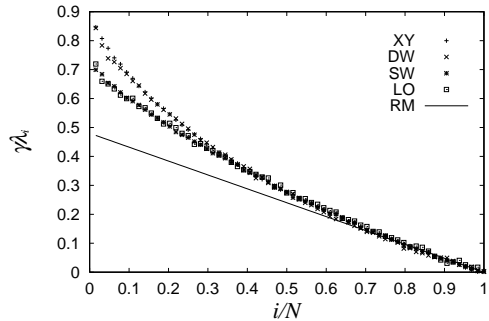}
    \caption{Universal behaviour of scaled Lyapunov spectra of four 
      models. The Lyapunov spectra coincide well among the four
      models in a range of large $i/N$, namely $0.4 \Lneq i/N \leq 1$,
      and they are not approximated by the straight line,
      which approximates them only in the range of $0.7 \Lneq i/N \leq
      1$. 
      The straight line is obtained with random matrices.
      Scale factor $\gamma$ is arranged in table
      \ref{tab:for-fig:comparison}.}
    \end{center}
    \label{fig:comparison}
\end{figure}

Next we show the invariant forms of Lyapunov spectra of the four
models together in figure \ref{fig:comparison}.
Scale factor $\gamma$ is arranged in table
\ref{tab:for-fig:comparison}.
The four spectra in figure \ref{fig:comparison} coincide in a
range of large $i/N$ ($0.4 \Lneq i/N \leq 1$),
while the straight solid line, which is obtained with random matrices, 
approximates them in a narrower range ($0.7 \Lneq i/N \leq 1$).
That is, the invariant forms do not depend on details of models
and are not straight,
and consequently, Lyapunov spectra of the four models have  
universality which is different from the one obtained in fully chaotic
systems. 

\begin{figure}
  \begin{center}
    \leavevmode
    \epsfxsize=6.5cm
    \epsffile{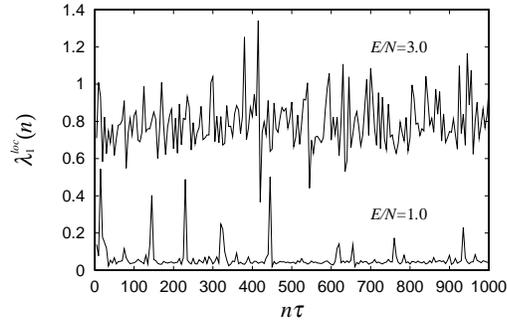}
    \caption{Temporal evolutions of local Lyapunov exponent
      $\lambda^{loc}_1 (n)$ in XY model. $N=10^3$.
      We set $\tau=10$. Lower and upper curves take
      $E/N=1.0$ and $3.0$ respectively,
      both of which yield the new universal Lyapunov spectra.
      Intermittency is found for $E/N=1.0$ but not for $E/N=3.0$.}
    \end{center}
    \label{fig:lle}
\end{figure}

Moreover, we show that the new universality appears even when systems
are not nearly integrable.
We confirm that systems are not nearly integrable 
through showing that KAM tori, many of which survive in nearly
integrable systems \cite{kam}, are not observed effectively.
We use the fact that linear instability is suppressed around KAM
tori since an orbit behaves like a regular one, while enhanced in
chaotic sea. 
Accordingly, in nearly integrable systems, intermittency of local
Lyapunov exponent $\lambda^{loc}_1(n)$ occurs which is defined as
follows.
\begin{equation}
  \lambda^{loc}_1 (n) = \frac{1}{\tau} 
  \int^{(n+1)\tau}_{n\tau} \lambda_1(t) dt,
\end{equation}
\begin{equation}
  \lambda_1(t) = \frac{\d}{\d t} \log |X(t)|,
\end{equation}
where $X(t)=(\delta q, \delta p)$ is a $2N$ dimensional tangent vector
following linearized Hamiltonian equations of motion,
equation (\ref{linearized-eq-motion}).
Figure \ref{fig:lle} shows two time series of local Lyapunov exponent
for $E/N=1.0$ and $3.0$ in XY model, both of which yield
the new universal Lyapunov spectra. Here $N=10^3$.
Intermittency is found for $E/N=1.0$ but not for $E/N=3.0$,
and hence existence of KAM tori does not seem to be related
to the new universality.
Consequently, the universality appears in systems with moderate
strength of chaos which are between nearly integrable and fully
chaotic. 

\section{3DFPU model and quadratic interactions}
\label{sec:FPU}

\begin{table}[hbtp]
  \begin{center}
    \leavevmode
    \caption{Behaviour of Lyapunov spectra in 3DFPU model.
      $k$ and $E/N$ are coupling constant and energy density,
      respectively. The sign S means straight behaviour of the
      Lyapunov spectrum, while the sign C curved behaviour. 
      The forms of curved Lyapunov spectra are in well agreement with
      the universal form.}
    \begin{tabular}{l | ccc}
      \br
      $k\setminus E/N$ & $1.0$ & $2.0$ & $3.0$ \\
      \mr
      $0.4$ & S & S & S \\
      $0.7$ & C & S & S \\
      $1.0$ & C & C & S \\
      \br
    \end{tabular}
    \label{tab:FPU}
  \end{center}
\end{table}

\begin{table}[hbtp]
  \begin{center}
    \leavevmode
    \caption{Ratios between time averages of the quadratic term $U_2$
      and the quartic terms $U_4$ of potential function in 3DFPU
      model.
      We subtract $1.50$ from each of the ratios to make a threshold
      clear,
      namely values arranged in this tables are
      $\overline{U}_2/\overline{U}_4 - 1.5$.
      The positive values are found at the places where
      the sign C appears in table \protect\ref{tab:FPU}.}
    \begin{tabular}{l | rrr}
      \br
      $k\setminus E/N$ & $1.0$ & $2.0$ & $3.0$ \\
      \mr
      $0.4$ & -0.60 & -0.91 & -1.02 \\
      $0.7$ & 0.33  & -0.33 & -0.62 \\
      $1.0$ & 1.55  &  0.37 & -0.09 \\
      \br
    \end{tabular}
    \label{tab:ratio}
  \end{center}
\end{table}

\begin{figure}
  \begin{center}
    \leavevmode
    \epsfxsize=6.5cm
    \epsffile{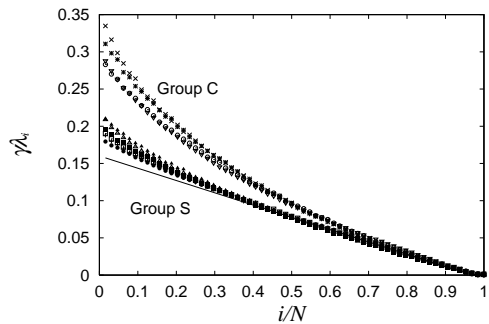}
    \caption{Lyapunov spectra in 3DFPU model.
      They are classified into group C and group S.
      Group C includes the universal spectrum which is in agreement
      with three spectra for 3DFPU,
      while group S consists of six spectra for 3DFPU which are
      straight.
      The straight line is a guide for eyes. 
      The classification is arranged in table \protect\ref{tab:FPU}
      with respect to $k$ and $E/N$.
      Scale factor $\gamma$ shown
      in table \protect\ref{tab:for-fig:fpu}.}
    \end{center}
    \label{fig:fpu}
\end{figure}

To probe what conditions produce the universality,
we study 3DFPU model with various values of energy density $E/N$
and coupling constant $k$ (see equation (\ref{hamiltonian-FPU})).
Lyapunov spectra for 3DFPU model are shown in figure \ref{fig:fpu}
with a Lyapunov spectrum for XY model belonging in the universality,
and they are classified into group C and group S.
Group C includes the universal Lyapunov spectrum which
is in well agreement with curved three spectra for 3DFPU,
and other spectra for 3DFPU are straight and belong to group S.
The classification is arranged in table \ref{tab:FPU}
with respect to $k$ and $E/N$.
Lyapunov spectra show curved forms when $k$ is large and $E/N$ is
low, 
we hence conjecture that quadratic terms of potential function are
dominant rather than quartic terms when the universality appears.

This conjecture is verified by taking ratios between time averages of
$U_2$ and $U_4$, namely $\overline{U_2}/\overline{U_4}$, where
\begin{equation}
  U_2 = \sum_{<ij>} \frac{k}{2} (q_i - q_j)^2,
\end{equation}
\begin{equation}
  U_4 = \sum_{<ij>} \frac{1}{4} (q_i - q_j)^4,
\end{equation}
and
\begin{equation}
  \overline{U}_n = \lim_{T\to\infty}\frac{1}{T}\int^T_0 dt U_n(t),
  \qquad (n=2,4).
\end{equation}
The ratios are arranged in table \ref{tab:ratio},
and they are large at the places where the sign C appears in table
\ref{tab:FPU}. 
Consequently, the conjecture is verified.

\section{Summary and discussions}
\label{sec:summary}


In order to consider structure of phase spaces in Hamiltonian
systems with moderate strength of chaos,
we numerically investigated Lyapunov spectra $\{\lambda_i\}$ for
five Hamiltonian systems with many degrees of freedom. 
We showed existence of universality of Lyapunov spectra,
which is defined as Lyapunov spectra approximately take the same form
regardless of energy and details of systems.
The universality gives a curved form for Lyapunov spectra,
and hence it is different from the one obtained in fully chaotic
systems. 

A feature of the new universality is that it appears in a finite
range of large $i/N$, where $N$ is degrees of freedom,
and accordingly, properties depending on energy or details of models
affect forms of Lyapunov spectra only in the range of small $i/N$
where is out of the universality.
In other words, we have only to focus on the range of small $i/N$
when we are interested in such individual properties.

We studied what conditions induce the universality
to understand the cause of it.
We showed that quadratic terms of a potential function dominate higher 
terms when the universality appears,
harmonic motions are therefore regarded as the base of global motions
in phase spaces.

We geometrically interpret the harmonic motions and that
the universality appears in a finite range of $i/N$ as follows.
Harmonic motions occur in high dimensional subspaces of phase spaces,
which correspond to the finite range of $i/N$,
and structure of phase space consists of chaotic sea and wrecks of
$n$-dimensional tori, where $n \leq N$ and $n$ may change for each
torus.
Here $n$-dimensional torus means, roughly speaking, direct product of 
$T^n$ and $(N-n)$-dimensional instability, which is hyperbolic or
complex.
Note $n=N$ is KAM torus.


We need further analyses to confirm whether this interpretation is
valid or not, and to understand the origin of the new
universality. 
We give two approaches which are geometrical and analytical.

A geometrical approach uses resonance of instability along an orbit.
Negative curvature of potential function induces positive Lyapunov
exponent. 
On the other hand, even the curvature is always positive, 
resonance of instability along a sample orbit gets the largest 
Lyapunov exponent to be positive \cite{pettini93}. 
Here let us assume this resonance theory can be extended not only to 
the largest Lyapunov exponent but to all the exponents. 
If the resonances are yielded by harmonic motions in high dimensional
subspace of phase space, 
then universality of Lyapunov spectra may be obtained because
harmonic motions are independent of details of models.

An analytical approach is to calculate decay rate spectrum of
harmonic motions. 
In a dissipative system, behaviour of Lyapunov spectrum
agrees with the decay rate spectrum of the linear fluctuation modes
from the stationary solution \cite{ikeda86}.
This suggests that to analyze decay rate spectrum is useful to
understand behaviour of Lyapunov spectrum. 
Hamiltonian systems are regarded as dissipative systems 
when we observe only subspaces of phase spaces,
and hence the decay rate spectrum must be useful because the new
universality appears only in high dimensional subspaces of phase
spaces.

Our goal is to understand global structure of phase spaces in
Hamiltonian systems with moderate strength of chaos. 
We must be allowed to come near the goal by understanding the cause of 
the new universality,
because ratios between Lyapunov exponents, which determine the form of 
the Lyapunov spectrum, seem to concern global structure of phase
spaces.
The new universality of Lyapunov spectra is an important clue to
reveal the global structure.

\ack
The author expresses grateful thanks to T.~Konishi for useful
discussions and careful reading of the manuscript.
The author thanks H.~Yamada, A.~Taruya and M.~Ishii for helpful
discussions. 
The author thanks the Computer Center of the Institute for
Molecular Science, for the use of the NEC SX-3/34R.


\pagebreak[4]
\appendix
\section{Tables of scale factor $\gamma$}

\begin{table}[hbtp]
  \begin{center}
    \leavevmode
    \caption{Scale factor $\gamma$ in figure \protect\ref{fig:scaled}.}
    \begin{tabular}{ cc | cc | cc | cc }
      \br
      \multicolumn{2}{c}{XY} & \multicolumn{2}{c}{DW} &
      \multicolumn{2}{c}{SW}& \multicolumn{2}{c}{LO} \\
      \mr
      $E/N$ & $\gamma$ & $E/N$ & $\gamma$ & 
      $E/N$ & $\gamma$ & $E/N$ & $\gamma$ \\
      \mr
      $1.0$ & $3.4$ & $1.0$ & $6.0$ & $1.0$ & $12$ & $1.0$ & $1.0$\\
      $2.0$ & $1.0$ & $1.4$ & $2.4$ & $3.0$ & $2.65$ & $3.0$ & $1.2$\\
      $3.0$ & $1/2.6$ & $1.8$ & $1.4$ & $5.0$ & $1.6$ & $5.0$ & $1.4$\\
      $4.0$ & $1/5.0$ & $2.2$ & $1.0$ & $7.0$ & $1.2$ & $7.0$ & $1.95$\\
      $5.0$ & $1/5.5$ & $100$ & $1/11$ & $9.0$ & $1.0$ &       & \\
      $100$ & $4.0$ &       &       &       &       &       & \\
      \br
    \end{tabular}
    \label{tab:for-fig:scaled}
  \end{center}
\end{table}

\begin{table}[hbtp]
  \begin{center}
    \leavevmode
    \caption{Rough estimations of ranges of $i/N$ where universal
      behaviour of Lyapunov spectra appears in four models.}
    \begin{tabular}{l | cccc }
      \br
      Model & XY & DW & SW & LO \\
      \mr
      $i/N$ & [0.4,1] & [0.31,1] & [0.16,1] & [0.11,1]\\
      \br
    \end{tabular}
    \label{tab:range}
  \end{center}
\end{table}

\begin{table}[hbtp]
  \begin{center}
    \leavevmode
    \caption{Scale factor $\gamma$ in figure
      \protect\ref{fig:comparison}.}
    \begin{tabular}{ c | cccc }
      \br
      Model    & XY    & DW    & SW    & LO \\
      \mr
      E/N      & $6.0$ & $2.0$ & $3.0$ & $3.0$ \\
      $\gamma$ & $1.0$ & $18$  & $12.7$ & $33$ \\
      \br
    \end{tabular}
    \label{tab:for-fig:comparison}
  \end{center}
\end{table}

\begin{table}[hbtp]
  \begin{center}
    \leavevmode
    \caption{Scale factor $\gamma$ in figure \protect\ref{fig:fpu}.}
    \begin{tabular}{c | lll}
      \br
      $k \setminus E/N$  & $1.0$ & $2.0$ & $3.0$ \\
      \mr
      0.4 & $1.0$ & $0.75$ & $0.66$ \\
      0.7 & $1.7$ & $0.91$ & $0.74$ \\
      1.0 & $2.25$ & $1.4$ & $0.92$ \\
      \br
    \end{tabular}
    \label{tab:for-fig:fpu}
  \end{center}
\end{table}

\end{document}